\documentstyle[12pt]{article}

\begin{document}
\def\tfrac#1#2{{\textstyle{#1\over #2}}}
\def\half{\tfrac{1}{2}}
\def\be{\begin{equation}}
\def\ee{\end{equation}}
\def\bea{\begin{eqnarray}}
\def\eea{\end{eqnarray}}

\def\eps{\varepsilon}
\def\mb{\mathbf}
\def\lo{\lambda_1}
\def\lt{\lambda_2}

\def\slr{$SL(2,R)$}
\def\so{$O(6,6)$}
\def\son{$O(6,6+n)$}
\def\su{$SU(1,1)$}
\def\tr{{\rm tr}}
\def\Tr{{\rm Tr}}

\hyphenation{trans-for-ma-tions}

\pagestyle{empty}
\rightline{UG-6/95}
\rightline{March 1995}
\rightline{hep-th/9503038}
\vspace{2truecm}
\centerline{\bf  Duality symmetry in four dimensional string actions}
\vspace{2truecm}
\centerline{\bf H.J.~Boonstra and M.~de Roo}
\vspace{.5truecm}
\centerline{Institute for Theoretical Physics}
\centerline{Nijenborgh 4, 9747 AG Groningen}
\centerline{The Netherlands}
\vspace{3truecm}
\centerline{ABSTRACT}

\vspace{.5truecm}

We reduce the dual version of $D=10$, $N=1$ supergravity
 coupled to $n$ vector fields to four dimensions, and
 derive the $SL(2,R)\times O(6,6+n)$ transformations which leave
 the equations of motion invariant. For $n=0$ \slr\ is also
 a symmetry of the action, but for $n>0$ only those \slr\
 transformations which act linearly on all fields leave the
 action invariant.
 The resulting four-dimensional
 theory is related to the bosonic part of
 the usual formulation of $N=4$ supergravity
 coupled to matter by a duality transformation.

\vfill\eject
\pagestyle{plain}

\noindent{\bf 1. Introduction}
\vspace{.3cm}

The symmetries of the low energy effective string action in four
 dimensions have received much attention recently because of their
 relation with $S$- and $T$-duality\footnote{For recent reviews,
 see, e.g,
 \cite{Giveon1,Sen1}.}.
 The toroidal compactification
 of the heterotic string can be studied by considering the
 reduction to four dimensions of $D=10$, $N=1$ supergravity
 coupled to $n=16$
\footnote{In the following, we will not restrict ourselves
 to this value of $n$.}
 abelian vector fields. This revealed
 an \son\  symmetry (related to $T$-duality)
 of the action \cite{Ma1}, as well as
 an \slr\ symmetry (related to $S$-duality)
 of the equations of motion \cite{Shap1}. Symmetries of the equations
 of motion of abelian vector fields involve a linear transformation
 of the Bianchi-identity, $\partial_\mu i\tilde F^{\mu\nu}$
 ($\tilde F^{\mu\nu}\equiv\half
  \epsilon^{\mu\nu\lambda\rho} F_{\lambda\rho}$),
 and the equation of motion, $\partial_\mu G^{\mu\nu}$, where
 $G \equiv -2 \partial{\cal L} / \partial F$. The linear
 transformation is of the form:
\be
      \left( \matrix{i\tilde F \cr G }\right)' =
             \omega \left( \matrix{i\tilde F \cr G }\right)\,.
\ee
The linear transformation must preserve the form of $G$ as obtained from
 the action, which restricts the possible $\omega$. For a single
 vector field the group of transformations can be
 $Sp(2,R)\simeq SL(2,R)$, if the
 vector field couples suitably to scalar fields which are also required to
 transform under \slr\ \cite{MK1}. In the absence of scalar fields
 the group of transformations is $O(2)$, which transforms electric and
 magnetic fields into each other.

The reduction of the {\it dual} version of $D=10$ supergravity
 has not been studied in as much detail from this point of view,
 in particular
 in the presence of additional (non)-abelian vector fields. In the
 dual version the two-index tensor gauge field is replaced
 by a six-index antisymmetric tensor gauge field.
 This is done by a standard duality transformation, which in $D$
 dimensions replaces an $m$-form gauge field by a $D-2-m$-form
 gauge field.
 In this paper we
 will discuss the symmetries of the four-dimensional theory which
 is obtained by reduction of this dual version to four dimensions.

 Our interest in this problem was raised by the work of Schwarz and
 Sen \cite{SS1}, who established that in the absence of additional
 vector fields
 in ten dimensions the resulting four-dimensional theory
 exhibits an \slr\  symmetry of the action.
 This result seems surprising from the point
 of view of $D=4$, $N=4$ supergravity theories, in which \slr\ appears
 only as a symmetry of the equations of motion\footnote{In $D=4$, $N=4$
 supergravity coupled to $n$ abelian vector multiplets
 \cite{DeRoo1}
 \son\ is a
 symmetry of the action, and \su\ (isomorphic to \slr) a symmetry of
 the equations of motion.}. On the other hand, having \slr\ as a symmetry
 of the action fits well in view of the conjectured
 string-fivebrane duality in ten
 dimensions \cite{Duff1,Strom1}. String-fivebrane duality
 is thought to interchange
 $T$- and $S$-duality \cite{SS1},
 and therefore the fivebrane effective action
 is expected to have (in $D=4$) an \slr\ symmetry. The
 assumption that the
 fivebrane effective action is built on the dual version of $D=10$
 supergravity, then leads to the \slr\ symmetry of the $D=4$ action.

To see whether or not the \slr\ symmetry of the action survives the
 coupling to (abelian) gauge fields, we will  work out the reduction
 to four dimensions in the presence of vector fields, determine
 the resulting symmetries of the action and equations of motion, and
 elucidate the relation between this result and $N=4$ supergravity
 theories. Our main result is that for $n>0$ \slr\ becomes a
 symmetry of the equations of motion only, together with \son.

The dual version of ten-dimensional supergravity, including its matter
 coupling and the reduction to four dimensions, was first discussed by
 Chamseddine \cite{Cham1}. The emphasis in this work was on
 properties of the resulting scalar potential, in view of
 possible supersymmetry breaking. The reduction of the bosonic part of
 the ten-dimensional action {\it without} additional vector fields
 was done in \cite{SS1}. We give the reduction {\it with}
 vector fields in section 2, and obtain the symmetries of the resulting
 action and equations of motion in section 3.
 We draw some conclusions on the relation with $N=4$ supergravity in
 section 4.

\vspace{.3cm}

\noindent \textbf{2. Reduction from $D=10$ to $D=4$}

\vspace{.3cm}

The bosonic part of the ten-dimensional action in the notation
and conventions of \cite{Berg2} is
\begin{eqnarray}
\label{10D}
{\cal{L}}_{D=10}=\sqrt{|\mathbf{G}|}{\mathbf\Phi}^{-3}
  \{ -\half{\mb R}({\mb G})+\tfrac{3}{4} {\mb H}_{ABC}{\mb H}^{ABC}
  +\tfrac{9}{2}({\mb\Phi}^{-1}\partial_M{\mb \Phi})^2 \nonumber\\
  +\beta\,\tr(-\tfrac{1}{4}{\mb F}^{MN}{\mb F}_{MN})\}
  -\tfrac{1}{4} i\beta\sqrt{2}\eps^{M_1 \cdots M_{10}}
  {\mb A}_{M_1 \cdots M_6}\tr({\mb F}_{M_7 M_8}{\mb F}_{M_9 M_{10}})\,,
\end{eqnarray}
where the field strength of the six-form gauge field $\mb{A}$ is
written in the form
\bea
{\mb H}_{ABC}=\tfrac{2}{3} i \sqrt{|{\mb G}^{-1}|}{\mb\Phi}^3
  \eps^{M_1 \cdots M_7}{}_{ABC}{\mb R}({\mb A})_{M_1 \cdots M_7}\,,\\
  {\mb R}({\mb A})_{M_1 \cdots M_7}\equiv\partial_{[M_1}{\mb A}_{M_2
  \cdots M_7]}\,.
\eea
The symbol tr denotes the trace in the adjoint representation,
and $\beta$ is the inverse squared of the Yang-Mills coupling
constant.
We use bold face capital letters to denote ten-dimensional fields.
Our index conventions are as follows:  $A,B,...$ are $10d$ flat indices,
$M,N,...$ are $10d$ curved indices, $m,n,...$ are $6d$ curved indices,
and $\mu,\nu,...$ are $4d$ curved indices.

For the reduction of (\ref{10D}) to $D=4$ we use the method of
 \cite{Scherk1}, which was applied to this case (without vector fields)
 in \cite{SS1}.
 The action has been divided in four parts, the Einstein-Hilbert action,
 the scalar kinetic
 part, the vector part and the scalar potential terms:
\bea\label{L4}
{\cal L}_{D=4}&=&{\cal L}_G + {\cal L}_S + {\cal L}_V + {\cal L}_P
\,,\\
\label{EH}
{\cal L}_G&=&-\tfrac{1}{2}\sqrt{|g|}\,R(g)\,,\\
\label{LS}
{\cal L}_S&=& -\sqrt{|g|}\,G^{mp}G^{nq}\{
  \tfrac{1}{8}(\partial_\mu G_{mn})(\partial^\mu G_{pq})
  +\tfrac{1}{8}(\partial_\mu B_{mn})(\partial^\mu B_{pq})
  \label{SK}\\
  &&\quad+\tfrac{\beta}{2}(\partial_\mu B_{mn})
  \,\tr\,(V_{[p} D^\mu V_{q]})
  +\tfrac{\beta^2}{2}\,\tr\,(V_{[m} D_\mu V_{n]})
  \,\tr\,(V_{[p} D^\mu V_{q]}) \}
  \nonumber\\
  &&\quad-\tfrac{\beta}{2}\sqrt{|g|}\,G^{mn}\,\tr\,(D_\mu V_m D^\mu V_n)
  -\tfrac{1}{8}\sqrt{|g|}\,\Tr\,[(\partial_\mu M)L(\partial^\mu
  M)L]\,,\nonumber\\
{\cal L}_V&=&-\tfrac{1}{8}\sqrt{|g|}\,G_{mn} {\cal F}_{\mu\nu}^m{}^T
  L^T M L {\cal F}^{\mu\nu\ n} - \tfrac{i}{16}\eps^{\mu\nu
  \rho\sigma}B_{mn}
  {\cal F}_{\mu\nu}^m{}^T L {\cal F}_{\rho\sigma}^n
  \label{LV}\\
  &&\quad+\beta\,\tr\,\{-\tfrac{1}{4}\sqrt{|g|}\,\lt (F_{\mu\nu}(V)+
  F_{\mu\nu}^m (A)V_m)^2
  \nonumber\\
  &&\quad-\tfrac{i}{8}\eps^{\mu\nu\rho\sigma}\lo
  F_{\mu\nu}(V) F_{\rho\sigma}(V)
  \nonumber\\
  &&\quad-\tfrac{i}{8}\eps^{\mu\nu\rho\sigma}(\lo F_{\mu\nu}^m (A)
  -F_{\mu\nu}^m (B))(F_{\rho\sigma}^n (A) V_m V_n
  +2F_{\rho\sigma}(V) V_m)\}\nonumber\,,\\
\label{LP}
{\cal L}_P&=&-\beta\sqrt{|g|}\,\lt^{-1}G^{mp}G^{nq}\{
  \tfrac{1}{4}\,\tr\,([V_m , V_n] [V_p , V_q])
  \label{SP}\\
  &&\quad+\tfrac{\beta}{6}G^{rs}\,\tr\,([V_m , V_n] V_r)
  \,\tr\,([V_p , V_q] V_s)\}\nonumber\,.
\eea

The origin of the four-dimensional fields is represented in Table 1.
 Note that the redefinition of $A^{m_1\ldots m_4}_{\mu_1\mu_2}$ to
 the scalars $B_{mn}$ involves a duality transformation. Of the
 other fields that arise from $\mathbf{A}$, the four-form field is
 completely absent from the four-dimensional action, while for the
 three-form field one can solve the equation of motion, which gives rise to
 the second term in the scalar potential ${\cal L}_P$ \cite{Cham1}.

The dilaton degree of freedom $\lt$ is contained in
 $\sqrt{G_{mn}}{\mb\Phi}^{-3}$. To obtain the form of the
 action (\ref{L4}-\ref{LP})
 a Weyl rescaling of the metric has been performed:
 $g_{\mu\nu}\rightarrow \lt^{-1}g_{\mu\nu}$.
 $D_\mu$ is the Yang-Mills covariant derivative.

\vspace{.25cm}
\begin{center}
\renewcommand{\arraystretch}{1.5}
\begin{tabular}{|l||l|l|l|}
\hline
\hfil$D=10$ &\hfil $D=4$\hfil&\hfil \# d.o.f\hfil & redefined to: \hfil\\
\hline
$\mathbf{G}$&$g_{\mu\nu}$&$2$&\hfil\\
\hfil&$A_{\mu}^m$&$12$&\hfil\\
\hfil&$G_{mn}$&$21$&\hfil\\
\hline
$\mathbf{A}$&$A^{m_1\ldots m_6}$&$1$&$\lambda_1$\\
\hfil&$A^{m_1\ldots m_5}_\mu$&$12$&$B_\mu^m$\\
\hfil&$A^{m_1\ldots m_4}_{\mu_1\mu_2}$&$15$&$B_{mn}$\\
\hfil&$A^{m_1\ldots m_3}_{\mu_1\ldots\mu_3}$&$0$&$-$\\
\hfil&$A^{m_1 m_2}_{\mu_1\ldots\mu_4}$&$0$&$-$\\
\hline
$\mathbf{\Phi}$&$\lambda_2$&$1$&\hfil\\
\hline
$\mathbf{V}$&$V_\mu$&2 dim $G$&\hfil\\
\hfil&$V_m$&6 dim $G$&\hfil\\
\hline
\end{tabular}
\renewcommand{\arraystretch}{1.0}
\end{center}
\vspace{.25truecm}
\noindent {\bf Table\ 1.} \ \ The four-dimensional fields
 and their ten-dimensional origin. The third column represents the
 number of degrees of freedom.
\vspace{.3truecm}

The kinetic terms of $A_\mu^m$ and $B_\mu^m$ in (\ref{LV}) have been
written in terms of the doublet
\bea
{\cal F}_{\mu\nu}^m\equiv
    \left(\matrix{
    F_{\mu\nu}^m (A) \cr
    F_{\mu\nu}^m (B) \cr}\right)\,.
\eea
In the absence of Yang-Mills fields ($\beta=0$),
the action is invariant under the \slr\
transforma- tions\footnote{Of course, the \slr\ symmetry is also
present in the action before the Weyl-rescaling,
 the redefinition leading to $B_{mn}$,
 and the elimination of $A^{m_1\ldots m_3}_{\mu_1\ldots\mu_3}$.}
\cite{SS1}
\bea\label{sl2}
{\cal F}\rightarrow \omega{\cal F}\,,\ \ \ \ M\rightarrow
\omega M\omega^T\,,\ \ \ \ \omega^T L\omega=L\,,
\eea
with the $2\times 2$ scalar matrix $M$ and \slr\  invariant
metric $L$ defined by
\bea
M={1\over \lt}\left(\matrix{
              1   & \lo            \cr
              \lo & \lo^2 + \lt^2  \cr}\right)\,,
\ \ \ \ \ L=\left(\matrix{ 0 & 1 \cr -1 & 0 \cr}\right)\,.
\eea
Note that the transformation of $M$ implies that $\lambda\equiv
 \lambda_1+i\lambda_2$ transforms under \slr\ as:
\be
\label{ltrans}
    \lambda\rightarrow {{c+d\lambda}\over {a+b\lambda}} \,,
 \ \ \ \ \ \
    \omega = \left( \matrix{ a&b \cr c&d } \right) \,,\qquad
    ad-bc =1\,.
\ee

In $D=4$ the pure $N=4$ supergravity multiplet has
 16, and an abelian vector multiplet 8 bosonic degrees of freedom.
 The reduction of $\mathbf{G},\ \mathbf{A}$ and $\mathbf{\Phi}$
 therefore gives the bosonic degrees of freedom of $N=4$
 supergravity coupled to  six  abelian vector multiplets.
 Each $\mathbf{V}$ in $D=10$ gives an additional vector multiplet
 in $D=4$.

The reduction of the two-index formulation of the $D=10$  effective
 action involves the two-form field $\mathbf{B}$ in place of the field
 $\mathbf{A}$. The result in $D=4$ involves scalars $B_{mn}$,
 vectors $B_\mu^m$, and the four-dimensional two-form field $B_{\mu\nu}$.
 The last gives only one degree of freedom in $D=4$, and can be
 replaced by a single scalar field by means of a duality transformation.
 This gives an action ${\cal L}_{D=4}'$, which has the same bosonic
 field content as the $N=4$ supergravity theories of \cite{DeRoo1}.

By performing these duality transformations in $D=4$ we are essentially
 undoing the $D=10$ duality transformation that relates the
 two-index to the six-index version. Indeed, a last duality transformation
 on the vectors $B_\mu^m$ of ${\cal L}_{D=4}'$ gives us precisely
 the action (\ref{L4}).

For $\beta=0$, i.e. without Yang-Mills fields, the action
(\ref{L4}) was obtained by Schwarz and Sen \cite{SS1}
from an action with  a doubled number of vector fields,
after elimination of an \slr\  invariant half of them.

\vspace{.3cm}

\noindent\textbf{3. Invariances of the equations of motion}

\vspace{.3cm}

If $\beta\neq 0$, i.e. if we have extra vector fields $V_\mu$,
it is not difficult to see that the action
(\ref{L4}) is no longer \slr\  invariant.
This is true even if all vector fields are abelian.

However, in the case of abelian vector fields,
we know that the equations of motion still
must be invariant under \slr, since this equation
of motion symmetry is well-established in the original
formulation of $D=4$ $N=4$ supergravity coupled to
abelian vector fields \cite{DeRoo1}, and the equations
of motion must be the same in both formulations.
For non-abelian vector fields this \slr\
invariance of the equations of motion does not hold, and
from now on we will assume that all vector fields are abelian.

The part of the action which is not \slr\  invariant
is the part of ${\cal L}_V$ containing the $n$ abelian
vector fields $V_\mu$.
A convenient way of finding the equations of motion invariance
is to introduce the doublet \cite{MK1}
\bea
{\cal D}(V)_{\mu\nu}=\left(\matrix{
-{i\tilde F}_{\mu\nu}(V) \cr G_{\mu\nu}(V)\cr}\right)\,,
\ \ \ \ \ \beta\sqrt{|g|}\, G^{\mu\nu}(V)\equiv 2{\partial{\cal L}
\over\partial F_{\mu\nu}(V)}\,,
\eea
where ${\tilde F}_{\mu\nu}(V) = \tfrac{1}{2}\eps_{\mu\nu\rho\sigma}
  F^{\rho\sigma}(V)$.
Then we can write the vector part of the action (where we
ignore the invariant kinetic terms for $A_\mu^m$
and $B_\mu^m$) as
\bea
{\cal L}_V = -\beta\sqrt{|g|}\,\tr\,\{-\tfrac{1}{4} G^{\mu\nu}(V)
  F_{\mu\nu}(V) + \tfrac{1}{4} {\cal D}(V)^{\mu\nu}{}^T  L
  {\cal F}_{\mu\nu}^m V_m\}\,.
\eea
Any linear \slr\  transformation of ${\cal D}$
which leaves the definition of $G(V)$ invariant is a
symmetry of the equations of motion \cite{MK1}.
In this case, we have
\bea
G_{\mu\nu}(V)=-\lt F_{\mu\nu}(V)-i\lo {\tilde F}_{\mu\nu}
(V)+C_{\mu\nu}\,,\nonumber\\
  C_{\mu\nu}=
  -\lt F_{\mu\nu}^m (A) V_m
  -i\lo {\tilde F}_{\mu\nu}^m (A) V_m +
  i{\tilde F}_{\mu\nu}^m (B)V_m\,.
\eea
and there are no restrictions on the $2\times 2$ \slr\
transformation matrices. Thus, ${\cal D}\rightarrow\omega
{\cal D}$, together with the transformations (\ref{sl2}),
is an \slr\
invariance of the equations of motion.
Note that the subgroup of $SL(2,R)$ transformations
\bea
\omega=\left(\matrix{a&0\cr c &\tfrac{1}{a}\cr}\right)
\eea
is a symmetry of the action (up to a total derivative, in case of the
 parameter $c$). This subgroup is formed by the transformations
 which act linearly on the scalar fields.
The parameter $a$ corresponds to a rescaling
of the vector fields $V_\mu$ and $A_\mu$ by the factor $a$ and
of $B_\mu$ by $\tfrac{1}{a}$, together with $\lambda_i
\rightarrow \tfrac{1}{a^2}\lambda_i\,,\ i=1,2$. The parameter $c$
shifts $\lambda_1$ to $\lambda_1+c$, and also $B_\mu$ to
$cA_{\mu}+B_\mu$. The vector $V_\mu$ is inert under this transformation.

In the same way, it is possible to find additional invariances
of the equations of motion. It is well-known that in the original
formulation of $D=4$, $N=4$ supergravity coupled to
abelian vector fields, the action has an \son\  invariance,
where $n$ is the number of additional abelian vector fields
$V_\mu^I$.
In our dual formulation, we do not have this symmetry of the
action,
so it must be an invariance of the equations of motion.
The scalar part ${\cal L}_S$ of the action
is invariant, which can be
seen by introducing the symmetric \son\  matrix $N$
of scalars \cite{Ma1},
\bea
N&=& \left(\matrix{G^{-1} & G^{-1} (B+W) &
  \sqrt{2\beta}G^{-1} V\cr
  (-B+W)G^{-1} & (G-B+W)G^{-1} (G+B+W) & \sqrt{2\beta}
  (G-B+W)G^{-1} V \cr
  \sqrt{2\beta} V^T G^{-1} & \sqrt{2\beta} V^T G^{-1} (G+B+W)
  & I_{n\times n} + 2\beta V^T G^{-1} V \cr}\right)\,,
  \nonumber\\
\eta&=&\left(\matrix{0 & I_{6\times6}&0\cr
  I_{6\times6}&0&0\cr0&0&-I_{n\times n}\cr}\right)\,,
\eea
where $\eta$ is the \son\  invariant metric, $N^T \eta N =\eta$.
The symmetric $6\times6$ matrix $(W)_{mn}$ is defined by
$W=\beta VV^T$, where $V$ is the
$6\times n$ matrix of scalars $(V)_m{}^I=V_m^I$. Furthermore,
$B$ is the antisymmetric matrix with components
$(B)_{mn}=B_{mn}$, and $G$ is still the internal metric.
Then the scalar kinetic part of the action is just
\bea
{\cal L_S}=\tfrac{1}{16}\sqrt{|g|}\,\Tr\,[(\partial_\mu N)\eta
  (\partial^\mu N)\eta] -\tfrac{1}{8}\sqrt{|g|}\,\Tr\,[
  (\partial_\mu M)L(\partial^\mu M)L]\,,
\eea
which is an \son\  invariant under
\bea\label{sctr}
N\rightarrow\Omega N\Omega^T\,,\ \ \ \ M\rightarrow M\,,
\ \ \ \ \Omega^T \eta\Omega=\eta\,.
\eea
The vector part of the action is not invariant.
To show that the equations of motion are invariant, we write
the complete vector part of the Lagrangian as
\bea\label{LVO}
{\cal L}_V=-\tfrac{1}{4}\sqrt{|g|}\,G^{\mu\nu}_a F_{\mu\nu}^a\,,
\eea
where $\sqrt{|g|}\,G^{\mu\nu}_a=-2{\partial{\cal L}\over\partial
F_{\mu\nu}^a}$, and the sum over $a$ extends over all vector fields
$A_\mu^m\,,B_\mu^m$ and $V_\mu^I$ ($a=1,\ldots,N$ with $N=12+n$).
 From the definition of $G$ and the Lagrangian (\ref{LVO}),
we get $G=F{\partial G\over\partial F}$ which has the general
solution
\bea
G^a = A^a{}_b F^b + i C^a{}_b {\tilde F}^b\,.
\eea
The $N\times N$ matrices $A$ and $C$ can be read off from the
action (\ref{LV}), and they depend on all scalars of the theory.
Now the \son\  equation of motion invariance
must be realized on the $2N$-dimensional multiplet
${\cal D}=(i{\tilde F}^a,G^a)$ \cite{MK1}. Indeed,
using the known transformation rules (\ref{sctr}) for the scalars,
we find that the \son\  transformations are
consistent with the definition of $G^a$. As expected,
the $2N$-dimensional multiplet ${\cal D}$
transforms in a reducible representation which
decomposes into two fundamental representations.
The resulting \son\  equations of motion
invariance is then given by
\bea
\left(\matrix{i{\tilde F}(A) \cr 2G(B) \cr -i\tilde F(V)\cr}\right)
\rightarrow \Omega
\left(\matrix{i{\tilde F}(A) \cr 2G(B) \cr -i\tilde F(V)\cr}\right)
\,,\ \ \ \ \left(\matrix{i{\tilde F}(B)\cr -2G(A)\cr -2G(V)\cr}
\right)\rightarrow \Omega
\left(\matrix{i{\tilde F}(B)\cr -2G(A)\cr -2G(V)\cr}\right)\,,
\eea
together with the transformations (\ref{sctr}).
Finally we note that the subgroup of \son\ which acts linearly,
or by a constant shift, on the
scalars leaves the action invariant (up to a total derivative).
This linear subgroup obviously contains $GL(6)\times O(n)$.

We have only discussed the equations of motion of the vector fields.
 The invariance of the other equations of motion under
 $SL(2,R)\times O(6,6+n)$ is guaranteed by the general arguments
 given in \cite{MK1}.

\vspace{0.3cm}

\noindent{\bf 4. Conclusions}

\vspace{0.3cm}

In the reduction to four dimensions, $N=1$ supergravity in $D=10$
 gives twelve abelian vector fields in $D=4$, of which six belong
 to the $N=4$ supergravity multiplet, and the other
 six belong to vector multiplets coupled to supergravity. In this
 paper only the bosonic sector of this theory was considered.

The two versions of $D=10$ supergravity (without additonal matter)
 give rise to different
 results in $D=4$: the two-index version gives \slr\ as a
 symmetry of the equations of motion, \so\ as a symmetry of the
 action. In the six-index version however \slr\ is a symmetry of
 the action, and \so\ an equation of motion symmetry. In
 section 2 we found that the two actions in $D=4$ are related
 by a duality transformation.

In \cite{DeRoo2} it was explained how different versions of
 $N=4$ supergravity can be obtained by duality transformations.
 Basically, $N=4$ supergravity coupled to abelian matter contains
 {\it two} \su\ symmetries: one is an equation of motion symmetry
 of the vector multiplets, the other is the \su\ symmetry of the
 superconformal $N=4$ Weyl multiplet \cite{Berg3}. These two
 symmetries can be identified in the process of matter coupling,
 as was done in \cite{DeRoo1}, but it is also possible to allow
 other isomorphisms between these two groups, determined by an
 arbitrary, constant {\su}-element $C$. The equations of motion
  remain unchanged under this modification, but the symmetries
 of the action do change. When (part of the) remaining symmetries
 of the action are gauged, inequivalent theories with local $N=4$
 supersymmetry are obtained.

The mechanism discussed in \cite{DeRoo2} can be used to explain the
 difference between the two reductions from $D=10$. The reduction of
 the two-index version from $D=10$ gives the result of \cite{DeRoo1}
 (after a duality transformation of $B_{\mu\nu}$).
 The reduction of the six-index version  corresponds to a particular
 choice of the \su-isomorphism in \cite{DeRoo2}. If the six
 abelian vector multiplets are coupled to supergravity with the
 \su-element (in the notation of \cite{DeRoo2}):
\be
    C = \left(\begin{array}{cc}
            -i & 0 \\
             0 & i
        \end{array}\right)
\ee
then one indeed finds the action (\ref{L4}). Note that the scalar
 fields in $N=4$ supergravity parametrize the coset
 $SU(1,1)/U(1)\times O(6,6+n)/(O(6)\times O(6+n))$. In
 \cite{DeRoo1,DeRoo2} this is expressed by imposing suitable constraints on
 the scalar fields. It has perhaps not been
 sufficiently realized that dimensional reduction from $D=10$
 provides a general,  explicit solution of these constraints.
 This solution, and the
 fact that by a duality transformation a version of $N=4$
 supergravity can be obtained which has an \su\ symmetry of the action
 is also derived in recent preprints of Zinoviev et al. \cite{Zino1}
 without reference to $D=10$.

The fact that part of the
 \slr\ symmetry of the effective action is broken by the
 coupling to additional vector fields suggests that the dual version
 of $D=10$, $N=1$ supergravity coupled to matter in the standard way
 \cite{Cham1} cannot be the low-energy effective action of the
 fivebrane.  In \cite{SS1},
 a modification of the ten-dimensional action including
 abelian vector fields
 was suggested which leads to  an \slr\
 invariant action in $D=4$.
 It will be interesting to see whether imposing \slr\ symmetry
 along these or other lines
 is a useful tool to obtain further insight in the
 effective action of the fivebrane.

\vfill\eject
\vspace{.3cm}

\noindent {\bf Acknowledgements}
\vspace{.3truecm}

It is a pleasure to thank Eric Bergshoeff for useful discussions
 on all kinds of duality.
The work of H.J.B.~was
performed as part of the research program of the ``Stichting voor
Fundamenteel Onderzoek der Materie'' (FOM).

\end{document}